# Parity Conservation in Supersymmetric Vector-Like Theories [1]

Hitoshi NISHINO

*Department of Physics*
*University of Maryland*
*College Park, MD 20742-4111, USA*
E-Mail: nishino@nscpmail.physics.umd.edu

**Abstract**

We show that parity is conserved in vector-like supersymmetric theories, such as supersymmetric QCD with massive quarks with no cubic couplings among chiral multiplets, based on fermionic path-integrals, originally developed by Vafa and Witten. We also look into the effect of supersymmetric breaking through gluino masses, and see that the parity-conservation is intact also in this case. Our conclusion is valid, when only bosonic parity-breaking observable terms are considered in path-integrals like the original Vafa-Witten formulation.

[1]This work is supported in part by NSF grant # PHY-93-41926.

## 1. Introduction

Non-perturbative chiral symmetry breaking [1] is an important aspect for phenomenological model building based on vector-like theories, such as in composite particle models [2][3]. It has been proven that parity symmetry is conserved in (non-supersymmetric) vector-like QCD theories, and is not broken spontaneously even non-perturbatively [4][5]. This proof [4] is based on the evaluation of fermionic path-integral, which gives always non-negative vacuum energy after adding parity-breaking terms. However, the question has risen whether the Vafa-Witten constraint [4][5] for the non-supersymmetric case can be avoided in supersymmetric vector-like theories because of the new interactions among gluino-quark-squarks [6][7], and whether parity is broken like other global symmetries [8]. These particular interaction terms with scalar-dependence seem to be the main obstruction for the proof for the positive definiteness of the determinant in the fermionic path-integral [5][6][7]. Moreover, the results about gauge symmetry breaking for massless supersymmetric QCD, when the number of flavor $N_f$ is smaller than the number of colors: $N_f < N$ [9] also seem to suggest the parity-breaking in supersymmetric vector-like theories. Although these observations seem reasonable, it also seems to contradict with the other universal wisdom about supersymmetry that supersymmetric vacuum is stable when the Witten index $\mathrm{Tr}\,(-1)^F$ [10] is non-zero, *e.g.,* $\mathrm{Tr}\,(-1)^F = N$ for the gauge group $SU(N)$, and therefore the vacuum energy most probably stays zero with no parity breaking. It has been also recently point out [11] that supersymmetric QCD has condenstate-free phase with no gluino condenstate. In our present paper we reconsider this subtle problem of parity breaking, and give a proof for parity conservation in supersymmetric vector-like theories with massive quarks.

The model we deal with in this paper is a globally supersymmetric vector-like theory with massive chiral multiplets coupled to a non-Abelian vector multiplet with no cubic coupling among chiral multiplets. Our proof is based on three major assumptions: First one for the massiveness of all the quark chiral multiplets, the second one about the absence of the Yukawa-couplings among chiral multiplets, and the third one that we rely on the method in [4] for purely bosonic parity-breaking observables. Therefore our method does not cover the fermionic parity-breaking observables like Wilson fermions [12] treted in lattice QCD.[2] The massiveness of quarks are also important for non-perturbative conservation of supersymmetry, due to well-defined non-zero Witten index $\mathrm{Tr}\,(-1)^F$ in such cases [10]. This is because supersymmetry is conserved, only if the vacuum energy is zero. Therefore the non-perturbative breaking of supersymmetry would cause the shift of vacuum energy, causing the breaking of parity symmetry [4][5]. Interestingly, we will find that parity is conserved also for a vector-like supersymmetric theory, like non-supersymmetric vector-like theory. We also look into the effect of gluino masses, which will not disturb the main body of the proof for supersymmetric case, and therefore parity is also conserved in broken supersymmetric vector-like theories.

[2]For reviews for supersymmetric QCD on lattice, see, e.g., [13]

## 2. Review for Non-Supersymmetric Vector-Like Theory

We start with reviewing the parity conservation in non-supersymmetric case [4][5] first, in order also to elucidate our notation. Suppose the total lagrangian $\mathcal{L}(\lambda) \equiv \mathcal{L} - \lambda X$ with a parameter $\lambda$ is a generalization of the parity-conserving lagrangian $\mathcal{L}$, such as that of QCD, with a parity-non-conserving observable $X$, such as the $F\widetilde{F}$-term, with a real constant $\lambda$. If parity is broken in the vacuum and $\langle X \rangle \neq 0$, then the theory can choose a vacuum state in which $\lambda \langle X \rangle < 0$ due to the signature ambiguity of $\langle X \rangle$. Hence the vacuum energy $E(\lambda)$ can be lower than $E(0)$ of the parity-conserving vacuum energy: $E(\lambda \neq 0) < E(0)$.

However, an explicit evaluation of path-integral reveals that this would not happen, *i.e.,* there is no such vacuum whose energy is lower than that of the parity-conserving one [4]. Consider the path-integral in Euclidian space for the vacuum energy $E(\lambda)$:

$$e^{-VE(\lambda)} = \int [dA_a{}^I][d\chi][d\overline{\chi}][d\xi][d\overline{\xi}] \exp\left[ - \int d^4x \, (\mathcal{L} + i\lambda X) \right] \quad , \tag{2.1}$$

where $V$ is the Euclidian volume, $A_a{}^I$ is the gluon field. The indices ${\scriptstyle I,\, J,\, \cdots}$ are for adjoint representations of the gauge group $G$. For example for $G = SU(N)$, we have ${\scriptstyle I,\, J,\, \cdots\, =\, 1,\, 2,\, \cdots,\, N^2-1}$. In order to clarify basic constituents of our system, we use 2-component spinors in this paper. Since we are dealing with a vector-like theory, our two-component Weyl spinors $\chi_i$ and $\xi^i$ with the flavor indices ${\scriptstyle i,\, j,\, \cdots\, =\, 1,\, 2,\, \cdots,\, N}$ for quarks are in the conjugate representations to each other. The factor of $i$ in the $\lambda X$-term is due to the usual Wick rotation. We specify the lagrangian as

$$\begin{aligned} &\mathcal{L} = -\tfrac{1}{4}(F_{ab}{}^I)^2 + \mathcal{L}_{\rm F} \quad , \\ &\mathcal{L}_{\rm F} = +i(\overline{\chi}^{\dot\alpha i} \not{D}^\beta{}_{\dot\alpha} \chi_{\beta i}) + i(\xi^{\alpha i} \not{D}_\alpha{}^{\dot\beta} \overline{\xi}_{\dot\beta i}) + m_i{}^j (\overline{\chi}^{\dot\alpha i} \overline{\xi}_{\dot\alpha j}) + m_j{}^i (\xi^{\alpha j} \chi_{\alpha i}) \quad , \end{aligned} \tag{2.2}$$

where $m \equiv (m_i{}^j)$ is an $N \times N$ hermitian mass matrix: $(m_i{}^j)^* = m_j{}^i$, which can be arranged to have only positive eigenvalues. Since we are dealing in this paper only with a vector-like theory, the quark fermions $\chi_i$ and $\xi^i$ are in the representations conjugate to each other, *e.g.,* $\mathbf{N}$ and $\mathbf{N}^*$-representations of $SU(N)$, respectively. Accordingly, our covariant derivative $D_a$ contains the minimal coupling of the gauge field to these fermions. We are using the notations similar to that in [14], *e.g.,* we use the Minkowskian four-dimensional (4D) vector indices ${\scriptstyle a,\, b,\, \cdots\, =\, 0,\, 1,\, 2,\, 3}$, with the signature $(\eta_{ab}) = \text{diag.}\ (+,-,-,-)$, while ${\scriptstyle \alpha,\, \beta,\, \cdots\, =\, 1,\, 2}$ and ${\scriptstyle \dot\alpha,\, \dot\beta,\, \cdots\, =\, \dot 1,\, \dot 2}$ for the 2-component spinors. Other relevant relations are such as

$$\begin{aligned} &\not{D}_{\alpha\dot\beta} \equiv (\sigma^c)_{\alpha\dot\beta} D_c \quad , \quad \left[ (\sigma^c)_{\alpha\dot\beta} \right]^* = (\sigma^c)_{\beta\dot\alpha} \quad , \quad \psi^\alpha = C^{\alpha\beta} \psi_\beta \quad , \quad \overline{\psi}_{\dot\alpha} = \overline{\psi}^{\dot\beta} C_{\dot\beta\dot\alpha} \quad , \\ &(\psi^\alpha)^\dagger = +\overline{\psi}^{\dot\alpha} \quad , \quad (\psi_\alpha)^\dagger = -\overline{\psi}_{\dot\alpha} \quad , \quad (\psi_1^{\alpha_1} \cdots \psi_j^{\alpha_j} \overline{\chi}_1^{\dot\beta_1} \cdots \overline{\chi}_k^{\dot\beta_k})^\dagger = \chi_k^{\beta_k} \cdots \chi_1^{\beta_1} \overline{\psi}_j^{\dot\alpha_j} \cdots \overline{\psi}_1^{\dot\alpha_1} \quad , \\ &\partial_a^\dagger = -\partial_a \quad , \quad (C_{\alpha\beta}) = (C_{\dot\alpha\dot\beta}) = \begin{pmatrix} 0 & -i \\ +i & 0 \end{pmatrix} \quad . \end{aligned} \tag{2.3}$$

Based on these, it is easy to confirm the reality of each term in our lagrangian.

As was shown by Vafa-Witten in a vector-like theory [5], the fermionic space for the path-integral (2.1) can be a direct sum of the positive and negative eigenstates of the Dirac operator in a finite volume $V$. Let $\chi_{\alpha i\mu}$ and $\overline{\xi}_{\dot\alpha i\mu}$ correspond to such eigenstates [15]:

$$\not{D}^{\beta\dot\alpha}\chi_{\beta i\mu} = +\mu\,\overline{\xi}^{\dot\alpha}{}_{i\mu} \quad , \qquad \not{D}^{\alpha\dot\beta}\overline{\xi}_{\dot\beta i\mu} = +\mu\,\chi^{\alpha}{}_{i\mu} \quad . \tag{2.4}$$

We then easily see that the set of $\chi_{\alpha(-\mu)} \equiv \chi_{\alpha(+\mu)}$, $\overline{\xi}_{\dot\alpha i(-\mu)} \equiv -\overline{\xi}_{\dot\alpha i(+\mu)}$ corresponds to the eigenvalue $-\mu$:

$$\not{D}^{\beta\dot\alpha}\chi_{\beta i(-\mu)} = -\mu\,\overline{\xi}^{\dot\alpha}{}_{i(-\mu)} \quad , \qquad \not{D}^{\alpha\dot\beta}\overline{\xi}_{\dot\beta i(-\mu)} = -\mu\,\chi^{\alpha}{}_{i(-\mu)} \quad . \tag{2.5}$$

Therefore the whole fermionic space is not only a direct sum of positive and negative eigenstates, but also they are always paired up between $+\mu > 0$ and $-\mu < 0$.

Or equivalently, in terms of a four-component Dirac spinor

$$\psi \equiv (\psi_{\underline{\alpha}i}) \equiv \begin{pmatrix} \chi_{\alpha i} \\ \overline{\xi}_{\dot\alpha i} \end{pmatrix} \quad , \qquad \overline{\psi} \equiv (\overline{\psi}^{\underline{\alpha}i}) \equiv \begin{pmatrix} \overline{\chi}^{\dot\alpha i} \\ \xi^{\alpha i} \end{pmatrix} \quad , \tag{2.6}$$

with the four-component spinorial indices $\underline{\alpha} \equiv (\alpha,\dot\alpha)$, $\underline{\beta} \equiv (\beta,\dot\beta)$, $\cdots$, we have

$$\widetilde{\not{D}}\psi = \begin{pmatrix} O & \not{D}_{\alpha}{}^{\dot\beta} \\ \not{D}^{\beta}{}_{\dot\alpha} & O \end{pmatrix} \begin{pmatrix} \chi_{\beta i} \\ \overline{\xi}_{\dot\beta i} \end{pmatrix} \quad , \qquad \widetilde{\not{D}} \equiv \begin{pmatrix} 0 & \not{D}_{\alpha}{}^{\dot\beta} \\ \not{D}^{\beta}{}_{\dot\alpha} & 0 \end{pmatrix} \quad . \tag{2.7}$$

We can also introduce the $2N \times 2N$ mass matrix $\widetilde{m}$ and the usual $4 \times 4$ $\gamma_5$-matrix for the four-component notation by

$$\widetilde{m} \equiv \begin{pmatrix} m_i{}^j & 0 \\ 0 & m_i{}^j \end{pmatrix} \quad , \qquad \gamma_5 \equiv \begin{pmatrix} \delta_\alpha{}^\beta & 0 \\ 0 & -\delta_{\dot\alpha}{}^{\dot\beta} \end{pmatrix} \quad . \tag{2.8}$$

Note also that $\widetilde{m}$ is hermitian: $\widetilde{m}^\dagger = \widetilde{m}$. The eigenstates in (2.6) are much transparent now for $\widetilde{\not{D}}$ as

$$\widetilde{\not{D}}\psi_\mu = +\mu\psi_\mu \quad , \qquad \widetilde{\not{D}}\psi_{(-\mu)} = -\mu\psi_{(-\mu)} \quad , \tag{2.9}$$

The eigenstates $\psi_{(-\mu)}$ correspond to $\psi_{(-\mu)} \equiv \gamma_5\psi_\mu$, because $\gamma_5$ satisfies $\{\gamma_5, \widetilde{\not{D}}\} = 0$, and therefore

$$\widetilde{\not{D}}\psi_{(-\mu)} = \widetilde{\not{D}}(\gamma_5\psi_\mu) = -\gamma_5\widetilde{\not{D}}\psi_\mu = -\mu\gamma_5\psi_\mu = (-\mu)\psi_{(-\mu)} \quad . \tag{2.10}$$

Therefore the pairing between $\mu$ and $-\mu$ eigenstates in (2.6) is clear. Accordingly, the lagrangian $\mathcal{L}_{\rm F}$ is simply

$$\mathcal{L}_{\rm F} = \overline{\psi}(i\widetilde{\not{D}} + \widetilde{m})\psi \quad , \tag{2.11}$$

Our path-integral (2.1) is now

$$
\begin{aligned}
e^{-VE(\lambda)} &= \int [dA_a{}^I] \int [d\psi][d\overline{\psi}]\, e^{-\int d^4x\left(\mathcal{L}_{\rm F}+\mathcal{L}_{\rm B}+i\lambda X\right)} \\
&= \int [dA_a{}^I] \int [d\psi][d\overline{\psi}]\, e^{-\int d^4x\left(\mathcal{L}_{\rm B}+i\lambda X\right)} \exp\left[-\int d^4x \overline{\psi}(i\widetilde{\not{D}}+\widetilde{m})\psi\right] \\
&= \int [dA_a{}^I]\, I_{\rm F} \exp\left[-\int d^4x(\mathcal{L}_{\rm B}+i\lambda X)\right] \quad . \qquad (2.11)
\end{aligned}
$$

Here $I_{\rm F}$ is the fermionic determinant from the fermionic path-integral

$$
\begin{aligned}
I_{\rm F} &\equiv \int [d\psi][d\overline{\psi}] \exp\left[-\int d^4x\, \overline{\psi}\,(i\widetilde{\not{D}}+\widetilde{m})\,\psi\right] \\
&= {\rm Det}\,(i\widetilde{\not{D}}+\widetilde{m}) \\
&= \prod_\mu \det\,(i\mu I_{2N}+\widetilde{m}) \\
&= (\det \widetilde{m})^{n_{\rm L0}+n_{\rm R0}} \left[\prod_{\mu>0}\det\,(i\mu I_{2N}+\widetilde{m})\right]\left[\prod_{\mu<0}\det\,(i\mu I_{2N}+\widetilde{m})\right] \\
&= (\det \widetilde{m})^{2n_0} \left[\prod_{\mu>0}\det\,(i\mu I_{2N}+\widetilde{m})\right]\left[\prod_{\mu>0}\det\,(-\,i\mu I_{2N}+\widetilde{m})\right] \qquad (2.12a) \\
&= (\det \widetilde{m})^{2n_0} \prod_{\mu>0}\det\,(i\mu I_{2N}+\widetilde{m})\det\,(-\,i\mu I_{2N}+\widetilde{m}) \\
&= (\det \widetilde{m})^{2n_0} \prod_{\mu>0}\det\,(i\mu I_{2N}+\widetilde{m})\det\,(+\,i\mu I_{2N}+\widetilde{m})^\dagger \qquad (2.12b) \\
&= (\det \widetilde{m})^{2n_0} \prod_{\mu>0}\left[\det\,(i\mu I_{2N}+\widetilde{m})\right]\left[\det\,(+\,i\mu I_{2N}+\widetilde{m})\right]^* \\
&= (\det \widetilde{m})^{2n_0} \left|\prod_{\mu>0}\det\,(i\mu I_{2N}+\widetilde{m})\right|^2 > 0 \qquad (2.12c)
\end{aligned}
$$

Here $I_{2N}$ is an $2N\times 2N$ unit matrix, and in (2.12a), $n_0$ is the number of $\mu=0$ modes, satisfying $n_{\rm L0}=n_{\rm R0}\equiv n_0$. This is because we have to consider only the instanton number zero background $n_{\rm L0}-n_{\rm R0}=0$ that is connected with the original vacuum with energy $E(\lambda=0)$ [16]. The previously-mentioned parings $\mu\leftrightarrow-\mu$ are used also in (2.12a). The determinants in (2.12a) are taken for the $2N\times 2N$ matrix for flavour indices, distinguished from the symbol 'Det' for the fermionic path-integral. In (2.12b) we have also used the hermiticity of $\widetilde{m}$. Now the positive definiteness of $I_{\rm F}$ is clear from (2.12). For the case of $N=1$, eq. (2.12c) is in agreement with [4].

Once the fermionic determinant (2.12) is positive, we see that the path-integral (2.1) is positive, except for the phase factor $\exp\,(i\lambda\int d^4xX)$, which does not lower the ground state energy. This is why $E(\lambda)$ must have a minimum only at $\lambda=0$ [4].

Before ending this section, we give the following lemma which will be of importance in the next section. Note that the lagraigian $\mathcal{L}_{\rm F}$ is rewritten as

$$\mathcal{L}_{\rm F} = \sum_\mu \mathcal{L}_\mu \equiv \sum_\mu \Big[ +i\mu(\overline{\chi}^{\dot{\alpha}i}{}_\mu \overline{\xi}_{\dot{\alpha}i\mu}) + i\mu(\xi^{\alpha i}{}_\mu \chi_{\alpha i\mu}) + m_i{}^j(\overline{\chi}^{\dot{\alpha}i}{}_\mu \overline{\xi}_{\dot{\alpha}j\mu}) + m_i{}^j(\xi^{\alpha i}{}_\mu \chi_{\alpha j\mu}) \Big] \ . \quad (2.13)$$

Accordingly, the path-integral (2.12) is also equivlent to

$$\begin{aligned} I_{\rm F} \ &= \left( \prod_\mu \int [d\chi_\mu][d\overline{\chi}_\mu][d\xi_\mu][d\overline{\xi}_\mu] \right) \\ &\qquad\qquad \times \exp\Big[ \int d^4x \sum_\mu \Big\{ \xi_{\alpha\mu} C^{\alpha\beta}(m + i\mu I_N)\chi_{\beta\mu} + \overline{\chi}_{\dot{\alpha}\mu} C^{\dot{\alpha}\dot{\beta}}(m + i\mu I_N)\overline{\xi}_{\dot{\beta}\mu} \Big\} \Big] \\ &= (\det \widetilde{m})^{2n_0} \left| \prod_{\mu>0} \det\left(i\mu I_{2N} + \widetilde{m}\right) \right|^2 > 0 \ . \end{aligned} \quad (2.14)$$

In other words, $I_{\rm F}$ in (2.12) can be computed in terms of 2-component spinors in (2.6). This relatinship will be helpful when we consider complicated mixed lagrangians between the gaugini and the quark/lepton fields in the next section.

## 3. Vector-Like Theory with Supersymmetry

We now generalize the above method to supersymmetric theories with no Yukawa couplings among quarks. Suppose we have the massive quark fermions $\chi_i$, $\overline{\chi}^i$, $\xi^i$, $\overline{\xi}_i$ together with the massless gluino Majorana fields $\lambda_\alpha{}^I$, $\overline{\lambda}_{\dot{\alpha}}{}^I$. As before, the indices $I, J, \cdots = 1, 2, \cdots, g = \dim G$ are for the adjoint representations of the gauge group $G$. All the fermion-dependent terms in our lagrangian are

$$\begin{aligned} \mathcal{L}_{\rm F} = \ & + i(\overline{\chi}^{\dot{\alpha}\,i} \not{D}^\beta{}_{\dot{\alpha}} \chi_{\beta i}) + i(\xi^{\alpha i} \not{D}_\alpha{}^{\dot{\beta}} \overline{\xi}_{\dot{\beta} i}) + i(\overline{\lambda}^{\dot{\alpha} I} \not{D}^\beta{}_{\dot{\alpha}} \lambda_\beta{}^I) + m_i{}^j(\overline{\chi}^{\dot{\alpha} i}\overline{\xi}_{\dot{\alpha} j}) + m_j{}^i(\xi^{\alpha j}\chi_{\alpha i}) \\ & + i(T^I)_i{}^j \Big[ z^{*i}(\lambda^{\alpha I}\chi_{\alpha j}) - z_j(\overline{\lambda}^{\dot{\alpha} I}\overline{\chi}_{\dot{\alpha}}{}^j) \Big] - i(T^I)_i{}^j \Big[ u^i(\overline{\lambda}^{\dot{\alpha} I}\overline{\xi}_{\dot{\alpha} j}) - u^*_j(\lambda^{\alpha I}\xi_\alpha{}^i) \Big] \ . \end{aligned} \quad (3.1)$$

The $z_i$ and $u^i$ are the spin $0$ fields (squarks) in the chiral multiplets $(z_i, \chi_i)$ and $(u^i, \xi^i)$ with $i, j, \cdots = 1, 2, \cdots, N$, in the representations $\mathbf{N}$ and $\mathbf{N}^*$. The $(T^I)_i{}^j$ are hermitian generators of the gauge group. As in (2.2), we can assume that $m_i{}^j$ is hermitian only with positive eigenvalues. The presence of these mixing terms with (pseudo)scalar-dependence have been considered to be the main obstruction for the parity-conservation in supersymmetric theories in the past [5][6][7], because they seem to prevent us from proving the positive definiteness of the fermionic determinant. However, we will see that this is not the obstruction. After the above prescription, there is no fermion-dependent term in the supersymmetric lagrangian $\mathcal{L}$ left over other than $\mathcal{L}_{\rm F}$: $\mathcal{L} = \mathcal{L}_{\rm F} + \mathcal{L}_{\rm B}$ with a purely bosonic lagrangian $\mathcal{L}_{\rm B}$.

We now consider the eigenstates for $\chi$ and $\xi$ as in (2.4), and rewrite all the $\chi$ and $\xi$-depenent terms in $\mathcal{L}_{\rm F}$, as

$$\mathcal{L}_{\chi,\xi} = \sum_\mu \mathcal{L}_{\chi,\xi,\mu} \equiv \sum_\mu \Big[ - \xi_\alpha{}^i{}_\mu C^{\alpha\beta} (m_i{}^j + i\mu\delta_i{}^j) \chi_{\beta j\mu} - \overline{\chi}_{\dot\alpha}{}^i{}_\mu C^{\dot\alpha\dot\beta} (m_i{}^j + i\mu\delta_i{}^j) \overline{\xi}_{\dot\beta j\mu}$$
$$- \xi_\alpha{}^i{}_\mu C^{\alpha\beta} \rho_{\beta i} - \overline{\chi}_{\dot\alpha}{}^i{}_\mu C^{\dot\alpha\dot\beta} \overline{\omega}_{\dot\beta i} - \overline{\xi}_{\dot\alpha i\mu} C^{\dot\alpha\dot\beta} \overline{\rho}_{\dot\beta}{}^i - \chi_{\alpha i\mu} C^{\alpha\beta} \omega_\beta{}^i \Big] ~, \quad (3.2)$$

where

$$\rho_{\alpha i} \equiv +i(T^I u^*)_i \lambda_\alpha{}^I ~, \qquad \omega_\alpha{}^i \equiv +i(z^* T^I)_i \lambda_\alpha{}^I ~,$$
$$\overline{\rho}_{\dot\alpha}{}^i \equiv -i(uT^I)^i \overline{\lambda}_{\dot\alpha}{}^I ~, \qquad \overline{\omega}_{\dot\alpha i} \equiv -i(T^I z)_i \overline{\lambda}_{\dot\alpha}{}^I ~. \quad (3.3)$$

As usual in path-integral, we can redefine the fields in such a way that the linear terms in $\chi$ or $\xi$ disappear. In our case, this can be done by the field redefinitions[3]

$$\widetilde{\xi}_\alpha{}^i{}_\mu \equiv \xi_\alpha{}^i{}_\mu + \omega_\alpha{}^j (m + i\mu I_N)^{-1}{}_j{}^i ~, \qquad \widetilde{\chi}_{\alpha i\mu} \equiv \chi_{\alpha i\mu} + (m + i\mu I_N)^{-1}{}_i{}^j \rho_{\alpha j} ~, \quad (3.4a)$$
$$\widetilde{\overline{\xi}}_{\dot\alpha i\mu} \equiv \overline{\xi}_{\dot\alpha i\mu} + (m + i\mu I_N)^{-1}{}_i{}^j \overline{\omega}_{\dot\alpha j} ~, \qquad \widetilde{\overline{\chi}}_{\dot\alpha}{}^i{}_\mu \equiv \overline{\chi}_{\dot\alpha}{}^i{}_\mu + \overline{\rho}_{\dot\alpha}{}^j (m + i\mu I_N)^{-1}{}_j{}^i ~, \quad (3.4b)$$

to have

$$\mathcal{L}_{\xi,\chi,\mu} = - \widetilde{\xi}_\alpha{}^i{}_\mu C^{\alpha\beta} (m + i\mu I_N)_i{}^j \widetilde{\chi}_{\beta j\mu} - \widetilde{\overline{\chi}}_{\dot\alpha}{}^i{}_\mu C^{\dot\alpha\dot\beta} (m + i\mu I_N)_i{}^j \widetilde{\overline{\xi}}_{\dot\beta j\mu}$$
$$+ \omega_\alpha{}^i C^{\alpha\beta} (m + i\mu I_N)^{-1}{}_i{}^j \rho_{\beta j} + \overline{\rho}_{\dot\alpha}{}^i C^{\dot\alpha\dot\beta} (m + i\mu I_N)^{-1}{}_i{}^j \overline{\omega}_{\dot\beta j} ~. \quad (3.5)$$

Since $m$ is hermitian only with positive eigenvalues, $m + i\mu I_N$ is also diagonalizable only with non-zero eigenvalues, and there is no problem for defining the inverse $(m + i\mu I_N)^{-1}$. After this, $\mathcal{L}_{\rm F}$ is now

$$\mathcal{L}_{\rm F} = \sum_\mu \mathcal{L}'_{\chi,\xi,\mu} + \sum_\mu \mathcal{L}_{\lambda^2,\mu} + \mathcal{L}_{\lambda \not{D} \lambda} ~, \quad (3.6)$$

where $\mathcal{L}'_{\chi,\xi,\mu}$ is the first line of (3.5) which coincides with the non-supersymmetric case (2.7), $\mathcal{L}_{\lambda\not{D}\lambda}$ is the gluino kinetic term, while $\mathcal{L}_{\lambda^2,\mu}$ is the $\lambda^2$ and $\overline{\lambda}^2$-terms after the field redefinition (3.4). Let us collect all of these $\lambda$-dependent terms into $\mathcal{L}_\lambda$:

$$\mathcal{L}_\lambda \equiv -i\overline{\lambda}_{\dot\alpha}{}^I \not{D}^{\beta\dot\alpha} \lambda_\beta{}^I - \sum_\mu \lambda_\alpha{}^I C^{\alpha\beta} M_\mu{}^{IJ} \lambda_\beta{}^J - \sum_\mu \overline{\lambda}_{\dot\alpha}{}^I C^{\dot\alpha\dot\beta} \overline{M}_\mu{}^{IJ} \overline{\lambda}_{\dot\beta}{}^J$$
$$\equiv -i\overline{\lambda}_{\dot\alpha}{}^I \not{D}^{\beta\dot\alpha} \lambda_\beta{}^I - \lambda_\alpha{}^I C^{\alpha\beta} M^{IJ} \lambda_\beta{}^J - \overline{\lambda}_{\dot\alpha}{}^I C^{\dot\alpha\dot\beta} \overline{M}^{IJ} \overline{\lambda}_{\dot\beta}{}^J ~, \quad (3.7)$$

where the matrices $M_\mu \equiv (M_\mu{}^{IJ})$, $\overline{M}_\mu \equiv (M_\mu{}^{IJ})$, $M \equiv (M^{IJ})$ and $\overline{M} \equiv (\overline{M}^{IJ})$ are defined by

$$M_\mu{}^{IJ} \equiv -(z^* \, T^{(I|} \, \widetilde{m}_\mu^{-1} T^{|J)} \, u^*) ~, \qquad \overline{M}_\mu{}^{IJ} \equiv -(u \, T^{(I|} \, \widetilde{m}_\mu^{-1} T^{|J)} \, z) ~,$$
$$M^{IJ} \equiv \sum_\mu M_\mu{}^{IJ} ~, \qquad \overline{M}^{IJ} \equiv \sum_\mu \overline{M}_\mu{}^{IJ} ~, \qquad \widetilde{m}_\mu \equiv m + i\mu I_N ~. \quad (3.8)$$

[3]Note that (3.4b) is not necessarily the hermitian conjugate of (3.4a). This is related to the hermiticity only by the combination of $+\mu$ and $-\mu$.

Due to the antisymmetry of $C^{\alpha\beta}$ and $C^{\dot\alpha\dot\beta}$, the matrices $M_\mu$, $\overline{M}_\mu$, $M$ and $\overline{M}$ are all symmetric in $I \leftrightarrow J$. Note also that

$$M_\mu{}^\dagger = M_\mu{}^* = \overline{M}_{-\mu} \quad , \qquad M^\dagger = M^* = \overline{M} \quad , \tag{3.9}$$

the latter of which is confirmed by the former under $\sum_\mu$, which is symmetric between $+\mu \leftrightarrow -\mu$. The $M$ is not necessarily hermitian, and it has both real and imaginary part. Now $\mathcal{L}_\lambda$ is rewritten as

$$\mathcal{L}_\lambda = -(\lambda^{\alpha I}, \overline{\lambda}^{\dot\alpha I}) \begin{pmatrix} \delta_\alpha{}^\beta M^{IJ} & i\delta^{IJ} \not{D}_\alpha{}^{\dot\beta} \\ i\delta^{IJ} \not{D}^\beta{}_{\dot\alpha} & \delta_{\dot\alpha}{}^{\dot\beta} \overline{M}^{IJ} \end{pmatrix} \begin{pmatrix} \lambda_\beta{}^J \\ \overline{\lambda}_{\dot\beta}{}^J \end{pmatrix} = \overline{\Lambda}\,(i\not{\mathcal{D}} + \mathcal{M})\Lambda \quad , \tag{3.10a}$$

$$\not{\mathcal{D}} \equiv \begin{pmatrix} O & \not{D}_\alpha{}^{\dot\beta} \\ \not{D}^\beta{}_{\dot\alpha} & O \end{pmatrix} \otimes I_g \quad , \qquad \mathcal{M} \equiv I_2 \otimes \begin{pmatrix} M & O \\ O & \overline{M} \end{pmatrix} \quad , \tag{3.10b}$$

$$\overline{\Lambda} \equiv -(\lambda^\alpha, \overline{\lambda}^{\dot\alpha}) = (\lambda_\beta, \overline{\lambda}_{\dot\beta}) \begin{pmatrix} C^{\beta\alpha} & O \\ O & C^{\dot\beta\dot\alpha} \end{pmatrix} = \Lambda^T \mathcal{C} \quad , \qquad \mathcal{C} \equiv \begin{pmatrix} C^{\alpha\beta} & O \\ O & C^{\dot\alpha\dot\beta} \end{pmatrix} \quad . \tag{3.10c}$$

The original fermionic lagrangian now is $\mathcal{L}_{\rm F} = \sum_\mu \mathcal{L}'_{\chi,\xi,\mu} + \mathcal{L}_\lambda$, and the total path-integral to be considered in the Euclidian space is

$$e^{-VE(\lambda)} = \int [dA_a{}^I][dz][dz^*][du][du^*] \; I_{\rm F} \; e^{-\int d^4x \left(\mathcal{L}_{\rm B} + i\lambda X\right)} \quad , \tag{3.11}$$

where the fermionic path-integral $I_{\rm F}$ is from (3.5) and (3.10) with $\mathcal{M}$ replaced by $\mathcal{M}$:

$$\begin{aligned} I_{\rm F} &= \int [d\chi][d\overline{\chi}][d\xi][d\overline{\xi}][d\lambda][d\overline{\lambda}] \exp\Big[ -\int d^4x \Big( \sum_\mu \mathcal{L}'_{\chi,\xi,\mu} + \mathcal{L}_\lambda \Big) \Big] \\ &= \Big( \prod_\mu \int [d\widetilde{\chi}_\mu][d\widetilde{\overline{\chi}}_\mu][d\widetilde{\xi}_\mu][d\widetilde{\overline{\xi}}_\mu] \, e^{-\int d^4x \, \mathcal{L}'_{\chi,\xi,\mu}} \Big) \int [d\lambda][d\overline{\lambda}] \, e^{-\int d^4x \, \mathcal{L}_\lambda} \\ &= (\det m)^{2n_0} \Big| \prod_{\mu>0} \det(m + i\mu I_N) \Big|^2 \, \big[ \mathrm{Det}\,(i\not{\mathcal{D}} + \mathcal{M}) \big]^{1/2} \quad . \end{aligned} \tag{3.12}$$

The first two factors are from the $\int [d\widetilde{\chi}][d\widetilde{\overline{\chi}}][d\widetilde{\xi}][d\widetilde{\overline{\xi}}]$-integral as in the non-supersymmetric case (2.14) now with the shifted variables $\widetilde{\chi}$, $\widetilde{\overline{\chi}}$, $\widetilde{\xi}$, $\widetilde{\overline{\xi}}$, and the remaining factor is from the $\int [d\lambda][d\overline{\lambda}]$-integral. There is potential phase ambiguity [17][18] for taking the square root in the last factor in (3.12). However, we will shortly show that there is no problem with this ambiguity in vector-like theories.

Even though we can not diagonalize $\not{\mathcal{D}}$ and $\mathcal{M}$ simultaneously, we still can use the eigenstate $|\nu\rangle$ for the eigenvalue $\nu \in \mathbb{R}$ of the operator $\not{\mathcal{D}}$:

$$\not{\mathcal{D}}\,|\nu\rangle = +\nu\,|\nu\rangle \quad , \qquad i.e., \quad \not{\mathcal{D}}\Lambda_\nu = \not{\mathcal{D}} \begin{pmatrix} \lambda_\nu \\ \overline{\lambda}_\nu \end{pmatrix} = +\nu \begin{pmatrix} \lambda_\nu \\ \overline{\lambda}_\nu \end{pmatrix} \equiv \nu\Lambda_\nu \quad , \tag{3.13}$$

where the subscript $_\nu$ on $\Lambda_\nu$ *etc.* denotes the eigenvalue for the four-component spinor $\Lambda$, with the adjoint index $^I$ omitted. As usual, we can define

$$\Gamma_5 \equiv I_2 \otimes \begin{pmatrix} I_g & O \\ O & -I_g \end{pmatrix} \quad , \tag{3.14}$$

satisfying $\{\Gamma_5, \not{\mathcal{D}}\} = 0$, so that an eigenstate $|-\nu\rangle$ of $\not{\mathcal{D}}$ can be constructed by

$$\not{\mathcal{D}}[\,\Gamma_5\,|\nu\rangle\,] == -\Gamma_5 \not{\mathcal{D}}\,|\nu\rangle = (-\nu)[\,\Gamma_5\,|\nu\rangle\,] \quad \Longrightarrow \quad \Gamma_5|\nu\rangle = |-\nu\rangle \ \ . \tag{3.15}$$

Therefore any eigenstate for $\forall\nu > 0$ is always paired up with an eigenstate $-\nu < 0$. We now see that a conjugate state $\langle\nu|$ is related to $|\nu\rangle$ as follows: Consider

$$\Lambda^\dagger = \begin{pmatrix} \lambda_\alpha \\ \overline{\lambda}_{\dot\alpha} \end{pmatrix}^\dagger = (-\,\overline{\lambda}_{\dot\alpha}\ ,\ -\lambda_\alpha) = (\lambda, \overline{\lambda}) \begin{pmatrix} O & -I_g \\ -I_g & O \end{pmatrix} = \Lambda^T \mathcal{F} = \overline{\Lambda} \mathcal{C}^{-1} \mathcal{F} \ \ , \tag{3.16a}$$

$$\mathcal{F} \equiv I_2 \otimes \begin{pmatrix} O & -I_g \\ -I_g & O \end{pmatrix} \quad , \quad \mathcal{F}^{-1} = \mathcal{F} \quad , \quad [\mathcal{F}, \mathcal{C}] = 0 \quad , \quad [\mathcal{F}, \mathcal{M}] = 0 \ \ . \tag{3.16b}$$

Here $\overline{\Lambda}$ is the usual Dirac conjugate of $\Lambda$, and $\mathcal{F}$ is needed for complex-conjugation. Therefore

$$|\nu\rangle^\dagger = \langle\nu|\,\mathcal{C}^{-1}\mathcal{F}^{-1} \quad , \quad \langle\nu|^{\,\dagger} = \mathcal{F}\mathcal{C}\,|\nu\rangle \ \ . \tag{3.17}$$

Accordingly, as in (3.15) we can confirm that

$$\langle -\nu| \, = \langle\nu|\,\Gamma_5 \ \ . \tag{3.18}$$

Other important relations needed are

$$\mathcal{F}\mathcal{M}\mathcal{F}^{-1} = \mathcal{M}^\dagger \quad , \quad \mathcal{C}\mathcal{M}\mathcal{C}^{-1} = \mathcal{M} \quad , \quad [\Gamma_5, \mathcal{M}] = 0 \ \ , \tag{3.19}$$

$$\mathcal{F}\not{\mathcal{D}}\mathcal{F}^{-1} = +\widehat{\not{\mathcal{D}}} \equiv \begin{pmatrix} 0 & \not{D}^{\beta}{}_{\dot\alpha} \\ \not{D}_{\alpha}{}^{\dot\beta} & 0 \end{pmatrix} \quad , \quad \mathcal{C}\not{\mathcal{D}}\mathcal{C}^{-1} = -\widehat{\not{\mathcal{D}}} \quad , \quad \not{\mathcal{D}}^\dagger = -\widehat{\not{\mathcal{D}}} \ \ , \tag{3.20}$$

as easily confirmed. Using these as well as (3.16), we get

$$\begin{aligned} \Gamma_5(i\not{\mathcal{D}} + \mathcal{M})\Gamma_5 = &\ -i\not{\mathcal{D}} + \mathcal{M} \\ = &\ +\mathcal{C}^{-1}\mathcal{F}^{-1}(i\widehat{\not{\mathcal{D}}} + \mathcal{M}^\dagger)\mathcal{F}\mathcal{C} = \mathcal{C}^{-1}\mathcal{F}^{-1}(-i\not{\mathcal{D}}^\dagger + \mathcal{M}^\dagger)\mathcal{F}\mathcal{C} \\ = &\ +\mathcal{C}^{-1}\mathcal{F}^{-1}(i\not{\mathcal{D}} + \mathcal{M})^\dagger\mathcal{F}\mathcal{C} \ \ . \end{aligned} \tag{3.21}$$

The determinant in the square root in the last factor in (3.12) can be re-expressed as the usual definition of the determinant in terms of exponential, trace and logarithmic functions:

$$\mathrm{Det}\,(i\not{\mathcal{D}} + \mathcal{M}) = (\det M)^{\widetilde{n}_0}(\det \overline{M})^{\widetilde{n}_0} \prod_{\nu\neq 0} \exp\,[\,\langle\nu|\,\ln(i\not{\mathcal{D}} + \mathcal{M})\,|\nu\rangle\,] \ \ . \tag{3.22}$$

As in the case of quarks, we consider only the instanton number zero background, so that the number of left- and right-handed zero-modes are the same: $\widetilde{n}_{\mathrm{L}0} = \widetilde{n}_{\mathrm{R}0} \equiv \widetilde{n}_0$, whose contributions in (3.22) can be computed separately, as

$$(\det M)^{\widetilde{n}_0}(\det \overline{M})^{\widetilde{n}_0} = (\det M)^{\widetilde{n}_0}(\det M)^{*\widetilde{n}_0} = |\det M\,|^{2\widetilde{n}_0} \ \ . \tag{3.23}$$

As for the $\nu \neq 0$ contributions, due to the pairing between the $|\nu\rangle$ and $|-\nu\rangle$, (3.22) is rewritten as

$$\begin{aligned}
&\mathrm{Det}\,(i\not{\mathcal{D}} + \mathcal{M}) \\
&= |\det M|^{2\widetilde{n}_0} \left[ \prod_{\nu>0} \exp\left[ \langle\nu|\ln(i\not{\mathcal{D}} + \mathcal{M})|\nu\rangle \right] \right] \left[ \prod_{\nu<0} \exp\left[ \langle\nu|\ln(i\not{\mathcal{D}} + \mathcal{M})|\nu\rangle \right] \right] \\
&= |\det M|^{2\widetilde{n}_0} \prod_{\nu>0} \exp\left[ \langle\nu|\ln(i\not{\mathcal{D}} + \mathcal{M})|\nu\rangle \right] \exp\left[ \langle-\nu|\ln(i\not{\mathcal{D}} + \mathcal{M})|-\nu\rangle \right] \ , \qquad (3.24)
\end{aligned}$$

where the exponent in the last factor is simplified by the aid of (3.17) and (3.21) as

$$\begin{aligned}
&\langle-\nu|\ln(i\not{\mathcal{D}} + \mathcal{M})|-\nu\rangle = \langle\nu|\Gamma_5 \ln(i\not{\mathcal{D}} + \mathcal{M})\Gamma_5|\nu\rangle \\
&= \langle\nu|\mathcal{C}^{-1}\mathcal{F}^{-1}\{\ln(i\not{\mathcal{D}} + \mathcal{M})\}^\dagger \mathcal{F}\mathcal{C}|\nu\rangle = \left[ \langle\nu|\ln(i\not{\mathcal{D}} + \mathcal{M})|\nu\rangle \right]^\dagger \\
&= \left[ \langle\nu|\ln(i\not{\mathcal{D}} + \mathcal{M})|\nu\rangle \right]^* \ . \qquad (3.25)
\end{aligned}$$

Therefore (3.24) is semi-positive definite:

$$\mathrm{Det}\,(i\not{\mathcal{D}} + \mathcal{M}) = |\det M|^{2\widetilde{n}_0} \prod_{\nu>0} \Big| \exp\langle\nu|\ln(i\not{\mathcal{D}} + \mathcal{M})|\nu\rangle \Big|^2 \geq 0 \ . \qquad (3.26)$$

Combining this with (3.12), we get the semi-positive definiteness of the fermionic determinant:

$$I_{\mathrm{F}} = |\det m|^{2n_0} |\det M|^{\widetilde{n}_0} \Big| \prod_{\mu>0} \det(m + i\mu I_N) \Big|^2 \prod_{\nu>0} \Big| \exp\langle\nu|\ln(i\not{\mathcal{D}} + \mathcal{M})|\nu\rangle \Big| \geq 0 \ . \quad (3.27)$$

Notice that the usual phase ambiguity when taking the square root [17][18] does not arise here, because of the semi-positive definite expression of (3.26), as conbributions from $|\nu\rangle$ and $|-\nu\rangle$ always in pairs. The main ingredient in this proof is the usage of the eigenstate $|\nu\rangle$ with the properties of $\Gamma_5$ and complex conjugations, which do not require the diagonalization of $\mathcal{M}$, or even its commutator with $\not{\mathcal{D}}$. The crucial procedure we have relied on is the expression of the determinant in terms of exponential, trace and logarithmic functions, which is to be the universal definition for a determinant.

Note that (3.27) implies only non-negativity of $I_{\mathrm{F}}$ which can still be zero. This is because the matrix $\mathcal{M}$ can depend on the scalar coordinates $Z \equiv (z, z^*, u, u^*)$. However, we can further show that there exists a measurable support (a connected domain with non-zero measure) in the $Z$-space, on which $I_{\mathrm{F}}(Z) > 0$ and non-zero. In fact, consider the particular point $Z_0 = 0$ on which $\mathcal{M} = O$ due to (3.8), (3.9) and (3.10). It follows that

$$I_{\mathrm{F}}(0) = |\det m|^{2n_0} |\det M|^{\widetilde{n}_0} \Big| \prod_{\mu>0} \det(m + i\mu I_N) \Big|^2 \prod_{\nu>0} \Big| \exp\langle\nu|\ln(i\nu)|\nu\rangle \Big| > 0 \ . \quad (3.28)$$

Once we get $I_{\mathrm{F}}(0) > 0$, then relying on the smoothness of $I_{\mathrm{F}}(Z)$ as a function of $Z$, we can conclude that $I_{\mathrm{F}}(Z) > 0$ on a measurable support including $Z_0 = 0$. The existence of a measurable support for $I_{\rm F} > 0$ leads us to the positivity of the path-integral measure $\int [dz][dz^*][du][du^*]\, I_{\rm F} > 0$, and we conclude that parity is conserved in supersymmetric vector-like theories.

In the above analysis, we have performed the most usual Wick rotation from the Minkowskian metric $(+,-,-,-)$ into the Euclidian one $(-,-,-,-)$ by replacing formally the coordinate $x^0 \rightarrow ix^4$. However, this may need more care, when it comes to the complex conjugation of spinors. Motivated by this, we have re-confirmed our result above by an alternative Wick rotation into the metric $(+,+,+,+)$. Additionally, the spinors in these Euclidian spaces are only 'formally' defined, in such a way that their complex conjugation rule is essentially the parallel to the Minkowskian case, like the simple replacement $x^0 \rightarrow ix^4$, and this is the very reason why the Feynman rules in the usual Euclidian path-integral is essentially the same as those in the Minkowskian. Rigorously speaking, spinors in the Euclidian spaces can exist only as $USp(2)$ spinors [19], and moreover the dotted and undotted spinors as eigenvectors of the $\gamma_5$-matrix are no longer related by complex conjugations [19].[4] For this precaution, we have also reformulated the Wick rotation, such that the dotted $\lambda_\alpha$ and undotted $\overline{\lambda}_{\dot\alpha}$ spinors in the final Euclidian space are not related to each other under complex conjugation, as they should be [19]. Interestingly, we have reached the same conclusion for the semi-positive definiteness of the determinant (3.27), even though the meaning of the bra- and cket-vectors are slightly modified, and all the pseudo-scalar Yukawa couplings with $\gamma_5$ acquire an extra factor of $i$ like the $\lambda X$-term in (2.1). One additional feature in this case we seem to rely on is that the gluino zero-modes are unstable and disappear from the physical spectrum, based on the analysis in ref. [21]. Since the details of this formulation is rather technical leading essentially to the same conclusion, we skip them in this paper.

## 4. Vector-Like Theory with Broken Supersymmetry

We mention the possibility of adding some gluino mass terms, which may be caused by some spontaneous, explicit, or non-perturbative breaking of supersymmetry. This is easily considered, by adding the gluino mass terms

$$\mathcal{L}_{m'\lambda^2} \equiv m'^{IJ}(\lambda^{\alpha I}\lambda_\alpha{}^J) + m'^{IJ}(\overline{\lambda}^{\dot\alpha I}\overline{\lambda}_{\dot\alpha}{}^J) \quad , \tag{4.1}$$

to our original lagrangian (3.1). Here $m' \equiv (m'^{IJ})$ is real and symmetric. Accordingly, (3.10a) is now

$$\mathcal{L}'_\lambda \equiv -(\lambda^\alpha, \overline{\lambda}^{\dot\alpha}) \begin{pmatrix} I_2 \otimes (M+m') & i\not{D} \otimes I_g \\ i\not{D}^T \otimes I_g & I_2 \otimes (\overline{M}+m') \end{pmatrix} \begin{pmatrix} \lambda_\beta \\ \overline{\lambda}_{\dot\beta} \end{pmatrix} \quad . \tag{4.2}$$

[4]This situation is similar to what is called Aiyah-Ward space-time with the signature $(+,+,-,-)$ studied in [20].

This implies that the matrix $M$ is replaced by $M+m'$ and $\overline{M}$ by $\overline{M}+m'$. Under this shift, the relations such as $M^\dagger = \overline{M}$ are intact. Eventually (3.27) is now replaced by

$$I'_{\rm F} = |\det m|^{2n_0} |\det M|^{\widetilde{n}_0} \left| \prod_{\mu>0} \det(m + i\mu I_N) \right|^2 \left| \prod_{\nu>0} \exp\langle\nu| \ln(i\not{\mathcal{D}} + \mathcal{M}') |\nu\rangle \right| \geq 0 \ , \qquad (4.3)$$

where $\mathcal{M}'$ is a $4g \times 4g$ matrix similar to $\mathcal{M}$ defined by

$$\mathcal{M}' \equiv \mathcal{M} + I_2 \otimes \begin{pmatrix} m' & O \\ O & m' \end{pmatrix} \quad , \qquad \mathcal{M}'^\dagger = \mathcal{M}'^* \quad . \qquad (4.4)$$

Hence the presence of $m'$ does not affect the semi-positive definiteness of the fermionic determinant. Accordingly, we can also show that $I'_{\rm F} > 0$ on a measurable support in the $Z$-space, and therefore we conclude that parity is conserved also in broken supersymmetric vector-like theories with non-zero gluino masses.

## 5. Concluding Remarks

In this paper we have shown the conservation of parity in supersymmetric vector-like theories. The main body of our proof is the confirmation that the determinant as the fermionic path-integral $I_{\rm F}$ is positive and non-zero on a measurable support under the bosonic integral $\int [dz][dz^*][du][du^*]$. We have also seen that the quark-gluino-squark mixing terms in the supersymmetric theory pose no problem. The supersymmetric vector-like theory seems to avoid the problem with these mixing terms thanks to parings between the eigenstates of the Dirac operator, despite of the complication caused by the mixing with gluini.

In our analysis of the fermionic path-integral, we first integrated over the quark fields $\chi$, $\overline{\chi}$, $\xi$, $\overline{\xi}$, making the computation more organized, instead of integrating over the gluino field first as in [6]. By so doing, we have seen that the final gluino path-integral is less involved and more controllable, in particular when we need to consider the Majorana gluino determinant which used to have subtlety with $\gamma_5$-pseudo-scalar couplings. We have understood that the usual phase ambiguity in the square root of the determinants for a 2-component spinor [17][18] does not arise in a vector-like theory, due to the pairing between the integrals over dotted and undotted spinors, combined with the pairing between the positive and negative eigenstates of the Dirac operator, including the zero-modes.

We have also studied the effect of the gluino masses, as a result of either spontaneous, explicit, or non-perturbative breaking of supersymmetry, and reached the conclusion that parity is also conserved in these cases with broken supersymmetry. In principle, we can also consider the squark masses caused by the supersymmetry breaking, but these terms are purely bosonic affecting only $\mathcal{L}_{\rm B}$, so that they are not expected to change our analysis or result in this paper.

Note that our result relies on the original method in [4], namely we deal only with bosonic parity-breaking observables in path-integrals. Therefore our result does not cover the parity-breaking *via* fermionic observables [12] which do not acquire the imaginary unit $i$ under the Wick-rotation.

Our result here seems to contradict with ref. [9], which shows that the vacuum structure is disturbed, when there are more colors than flavors: $N_f < N$. However, there is actually no conflict, because we interpret this as the result of the masslessness of quarks treated in ref. [9]. In our system, due to the massive quarks from the outset, the Witten index $\mathrm{Tr}\,(-1)^F$ is well-defined and non-zero [10], *e.g.*, $\mathrm{Tr}\,(-1)^F = N$ for the $SU(N)$ gauge group. Hence we expect no breaking of supersymmetry even at the non-perturbative level, which would have ruined the foundation of our proof. Since the topological stability due to the well-defined $\mathrm{Tr}\,(-1)^F$ is reliable for massive quarks, it is quite natural that chiral symmetry or parity symmetry is also conserved. Additionally, our parity-conservation is also consistent with the result of [11] about the phase with unbroken discrete axial symmetry. Some subtlety arises, when the masses of the quarks become zero, because in such a case the Witten index $\mathrm{Tr}\,(-1)^F$ is no longer well-defined [10][22], and therefore the vacuum loses its stability against chiral or parity breakings as in [9]. From this viewpoint, we see no contradiction of our result with refs. [9][22], in which the masses of quarks are zero from the outset.

We have seen that the supersymmetry breaking, if its only effect on fermions is the gluino masses, does not alter the conservation of parity. Even though this statement seems contradictory with the previous paragraph, we understand that the breaking of supersymmetry lifts the vacuum energy higher than the original supersymmetric and parity-conserving vacuum with $E(0) = 0$.

We can try to apply our method to other arbitrary global or discrete symmetries, such as baryon number, in a supersymmetric vector-like theory, using the prescription using upper bounds for fermionic propagators in ref. [23]. However, there seems to be an obstruction caused by the zero-ness of mass eigenvalues, *e.g.*, our matrix $\mathcal{M}$ in (3.10) hitting zeros, that upsets the upper bound for fermionic propagators, undermining the foundation for the stability of fermion-anti-fermions Greens functions against symmetry-breaking parameters [5][23]. In other words, our method using the semi-positive definite fermionic determinant is powerful only for parity symmetry, or other symmetries based only on the vacuum to vacuum amplitude.

We are grateful for J.C. Pati for bringing about the problem with parity conservation in supersymmetric QCD, and for other helpful discussions. We are also indebted to M. Luty, R. Mohapatra, Y. Shamir, and E. Witten for important suggestions and comments. Additional acknowledgement is due to M. Cvetič for a copy of reference [6], and other important discussions. Special acknowledgement is for C. Vafa for pointing out mistakes in an earlier version of the manuscript.